\documentclass{article}[12pt]
\usepackage[letterpaper,margin=1in]{geometry}
\usepackage{amscd,amsmath,amsthm,amsfonts,amssymb}
\usepackage{graphicx}
\usepackage{multicol}
\usepackage{color}
\numberwithin{equation}{section}
\newcommand{\comment}[1]{}
\newcommand{\disp}{\displaystyle}
\def\sech{{\rm sech}}
\def\cosh{{\rm cosh}}

\def\csch{{\rm csch}}

\def\Imag{{\rm Im}}

\def\PT{$\mathcal{PT}$}

\def\ri{{i}}

\def\rd{{\rm d}}

\def\re{{\rm e}}

\title{Exponential asymptotics for solitons in $\mathcal{PT}$-symmetric periodic potentials}

\begin{document}

\author{Sean D. Nixon and Jianke Yang \\  \\
Department of Mathematics and Statistics \\ University of Vermont
\\ Burlington, VT 05401, USA}
\date{ }
\maketitle

\begin{abstract}

Solitons in one-dimensional parity-time (\PT)-symmetric periodic
potentials are studied using exponential asymptotics. The new
feature of this exponential asymptotics is that, unlike conservative
periodic potentials, the inner and outer integral equations arising
in this analysis are both coupled systems due to complex-valued
solitons. Solving these coupled systems, we show that two soliton
families bifurcate out from each Bloch-band edge for either
self-focusing or self-defocusing nonlinearity. An asymptotic
expression for the eigenvalues associated with the linear stability
of these soliton families is also derived. This formula shows that
one of these two soliton families near band edges is always
unstable, while the other can be stable. In addition, infinite
families of \PT-symmetric multi-soliton bound states are constructed
by matching the exponentially small tails from two neighboring
solitons. These analytical predictions are compared with numerics.
Overall agreements are observed, and minor differences explained.

\end{abstract}
\bigskip

\section{Introduction}

Nonlinear propagation of waves in periodic media is of keen interest
in the fields of applied mathematics and physics \cite{Yang_SIAM},
with applications that range from nonlinear photonics
\cite{Kivshar2003,Skorobogatiy2009} to Bose-Einstein condensates
\cite{Morsch2006,Kevrekidis2008}. Recently this research has
overlapped with the study of parity-time (\PT) symmetry from quantum
mechanics. \PT-symmetric systems have the unintuitive property that
they can possess all-real linear spectra despite the presence of
gain and loss \cite{Bender1998,Muga2005,Christodoulides2007}. When
\PT-symmetry is coupled with nonlinearity, a new phenomenon is that
the nonlinear \PT system can support continuous families of
solitons, which is remarkable for non-conservative systems
\cite{Musslimani2008,Wang2011,Lu2011,Abdullaev2011,Nixon2012,Zeng2012,Li2012,Driben2011,Zezyulin2012,Moreira2012,Kevrekidis2013,Kartashov2013}.
These soliton families, however, have to be \PT-symmetric in
one-dimensional and most higher-dimensional cases, where
\PT-symmetry breaking of solitons is forbidden \cite{Yang2014}.

In this paper, we analytically study solitons and their linear
stability in the one-dimensional nonlinear Schr\"odinger (NLS)
equation with a \PT-symmetric periodic potential. We examine
small-amplitude solitons which bifurcate out from infinitesimal
Bloch modes taking the form of slowly varying Bloch-wave packets.
While the packet envelope can be readily found to satisfy the
familiar potential-free NLS equation and thus have a sech-shape, the
position of the envelope relative to the periodic potential is
harder to determine because it hinges on effects that are
exponentially small in the soliton amplitude.

For the case of strictly-real periodic potentials, the exponential
asymptotics method for analyzing low-amplitude solitons has been
developed before \cite{Hwang2011,Akylas2012,Hwang2012,Nixon2013}
(see also \cite{Akylas1997,Akylas2000} on the fifth-order
Korteweg-de Vries equation). In this method, the Fourier transform
is taken with respect to the slow spatial variable of the envelope
function, motivated by the fact that solitary-wave tails in the
physical domain are controlled by pole singularities near the real
axis of the wavenumber space. Residues of these poles, which are
exponentially small, are then calculated by matched asymptotics near
the poles and away from the poles. Upon inverting the Fourier
transform, these poles of exponentially small strength give rise to
growing tails of exponentially small amplitudes in the physical
solution. These growing tails turn out to be dependent on the
envelope position. Then demanding these growing tails to vanish
would yield the true envelope position of low-amplitude solitons.
Linear-stability eigenvalues of these low-amplitude solitons can
also be derived by utilizing the growing-tail formula, thus linear
stability of these solitons can be determined by exponential
asymptotics as well. In addition, by matching tails of several wave
packets, infinite families of multi-packet solitary waves (referred
to as multi-soliton bound states) can be constructed. In particular,
if the periodic potential is symmetric, then beside families of
symmetric bound states, families of asymmetric bound states also
exist through symmetry-breaking bifurcation.

In this paper, we develop the exponential asymptotics analysis for
low-amplitude solitons in complex \PT-symmetric periodic potentials.
Unlike the real-potential case, here the soliton is complex-valued,
thus solution behavior near the two nearest pole singularities in
the wavenumber domain is coupled. Furthermore, the solution behavior
away from the poles also depends on a coupled equation. Fortunately
these coupled equations can be solved, thus growing tails of
exponentially small amplitudes in Bloch-wave packets can still be
derived. This tail formula reveals the existence of two
low-amplitude solitons, with envelopes located at the point of \PT
symmetry and half-period away from it. Calculations of
linear-stability eigenvalues show that near band edges, one of these
two soliton families is always unstable, while the other family can
be stable. Two-soliton bound states are also derived by matching the
growing tail of one soliton to the decaying tail of a neighboring
soliton. Most of these analytical predictions are confirmed by our
direct numerical computations. The only exception is on
non-\PT-symmetric bound states, where leading-order tail matching of
exponential asymptotics predicts the existence of such bound states,
but numerical computations disprove their existence (this numerical
nonexistence is consistent with the earlier analysis in
\cite{Yang2014}). However, the numerical residue error of these
approximate non-\PT-symmetric bound states is found to be extremely
small, which suggests that the nonexistence of such soliton states
is due to higher-order effects of exponential asymptotics.

\section{$\mathcal{PT}$-symmetric solitons}

We consider the one-dimensional nonlinear Schr\"{o}dinger equation
with a periodic potential $V(x)$,
\begin{equation} \label{Eq:NLS}
\ri \Psi_t + \Psi_{xx} - V(x)\Psi + \sigma |\Psi |^2 \Psi = 0,
\end{equation}
where $V(x)$ is complex and satisfies the $\mathcal{PT}$-symmetry
condition $V(x) = V^*(-x)$, with the asterisk representing complex
conjugation, and $\sigma=\pm 1$ is the sign of nonlinearity.
Throughout this paper, the period of the potential $V(x)$ is taken
to be equal to $\pi$ without any loss of generality.

We search for soliton solutions of the form
\begin{equation}  \label{e:soliton}
\Psi(t,x) = \psi(x) \re^{-\ri \mu t},
\end{equation}
where $\mu$ is the propagation constant and $\psi$ is a complex amplitude function solving the equation
\begin{equation} \label{Eq:psi}
\psi_{xx} + \left( \mu - V \right) \psi + \sigma |\psi|^2\psi = 0.
\end{equation}

When $\psi$ is infinitesimal, equation \eqref{Eq:psi} reduces to the
linear Schr\"{o}dinger equation
\begin{equation*}
\psi_{xx} + \left( \mu - V \right) \psi = 0.
\end{equation*}
This equation, by the Bloch-Floquet Theorem, has bounded solutions of the form
\begin{equation*}
p(x; \mu)= \re^{\ri k x } \widetilde{p}(x; k),
\end{equation*}
where $\widetilde{p}(x; k)$ is periodic with the same period $\pi$
as the potential $V(x)$, $\mu=\mu(k)$ is the dispersion relation
which forms Bloch bands, and $k$ lies in the first Brillouin zone
$-1\le k\le 1$.

For \PT-symmetric periodic potentials, the Bloch bands can be
all-real. For instance, the \PT potential
\begin{equation} \label{Eq:ExLatt}
V(x) = V_0\left[\sin^2(x) + i W_0 \sin (2x)\right]
\end{equation}
has all-real Bloch bands when $V_0>0$ and $|W_0|\le 1/2$
\cite{Musslimani2008,Nixon2012}. But it should be recognized that
Bloch bands of a \PT-symmetric periodic potential can also be
complex. For instance, for the above potential (\ref{Eq:ExLatt}),
part of the Bloch bands becomes complex when $|W_0|>1/2$
\cite{Musslimani2008,Nixon2012}. When the Bloch bands are complex,
the corresponding linear Bloch modes are unstable. As a consequence,
any soliton solution (\ref{e:soliton}) in Eq. (\ref{Eq:NLS}) is
linearly unstable too.

The only assumption we make for the ensuing exponential asymptotics
analysis is that the Bloch band around a band edge $\mu=\mu_0$ is
real. Under this assumption, we analyze how low-amplitude solitary
waves bifurcate out from this band edge as $\mu$ moves into the band
gap.

Near the band edge, solutions to Eq. (\ref{Eq:psi}) are
low-amplitude Bloch-wave packets which can be expanded into
perturbation series
\begin{align} \label{Eq:psiSeries}
\psi(x, X) & = \epsilon \psi_0(x,X) + \epsilon^2 \psi_1(x,X) + \epsilon^3\psi_2(x,X) + O(\epsilon^4),
\\
\mu & = \mu_0 + \eta \epsilon^2,
\end{align}
where $X=\epsilon x$, $0< \epsilon \ll 1$, and $\eta=\pm 1$.
Substituting this expansion into equation \eqref{Eq:psi} yields
\begin{equation} \label{Eq:MultiScale}
L_0 \psi +  \epsilon 2\psi_{xX}  + \epsilon^2  \psi_{XX} + \sigma |\psi|^2\psi +  \eta \epsilon^2\psi = 0,
\end{equation}
where
\begin{equation*}
L_0 \equiv  \partial_{xx} + \mu_0 - V(x).
\end{equation*}
Substituting (\ref{Eq:psiSeries}) into (\ref{Eq:MultiScale}) and
performing standard perturbation calculations
\cite{Hwang2011,Pelinovsky2004}, we arrive at the solution
\begin{equation}  \label{e:PsixX}
\psi(x,X) = \epsilon A(X) p(x) + \epsilon^2 A'(X) \nu(x) + O(\epsilon^3),
\end{equation}
where $p(x)\equiv p(x; \mu_0)$ is the Bloch mode at band edge
$\mu_0$, $\nu(x)$ solves
\begin{align*}
L_0 \nu  &= - 2 p_x,
\end{align*}
the envelope function $A(X)$ solves
\begin{equation} \label{Eq:Envelope}
D A_{XX}+\eta A  + \sigma a A^3 = 0
\end{equation}
with
\begin{equation} \label{Eq:aD}
D = \frac{1}{2} \left. \frac{d^2\mu}{dk^2} \right|_{\mu=\mu_0},  \quad  a = \frac{\langle |p|^2 p, p^* \rangle}{ \langle p , p^* \rangle},
\end{equation}
and the inner product is defined as $\langle f,
g\rangle=\int_{-\pi}^{\pi} f(x) g^*(x) dx$. In deriving this $D$
formula, the identity
\begin{equation*}
\frac{\langle 2\nu_x + p , p^*\rangle}{ \langle p , p^*\rangle}=\frac{1}{2} \left. \frac{d^2\mu}{dk^2} \right|_{\mu=\mu_0}
\end{equation*}
has been used (a similar identity for conservative periodic
potentials has been reported before
\cite{Yang_SIAM,Pelinovsky2004}). To make the Bloch mode $p(x)$
unique, we scale it so that $p(0)=1$. To avoid ambiguity of the
homogeneous term in $\nu$, we impose the condition $\langle
\nu,p^*\rangle = 0$.

When sgn($\sigma$) = sgn($D$) = $-$sgn($\eta$), Eq.
\eqref{Eq:Envelope} admits a solitary wave solution
\begin{equation} \label{Eq:Asech}
A(X) = \alpha ~\sech \left(\frac{X-X_0}{\beta}\right),
\end{equation}
where the constants $\alpha$ and $\beta$ are defined as
\begin{equation*}
\alpha = \sqrt{2/a},~~~~\beta = \sqrt{|D|}.
\end{equation*}
Note that at each band edge $\mu_0$, sgn($\eta$) is chosen as
$-$sgn($D$), meaning that $\mu = \mu_0 + \eta \epsilon^2$ lies in
the band gap. In addition, sgn($\sigma$) is chosen as sgn($D$),
meaning that the soliton solution only exists for either the
focusing or defocusing nonlinearity. However, the soliton position
parameter $X_0$ is free since the envelope equation
(\ref{Eq:Envelope}) is translation-invariant.

\begin{figure}[htbp!]
\begin{center}
\includegraphics[width=4.5in]{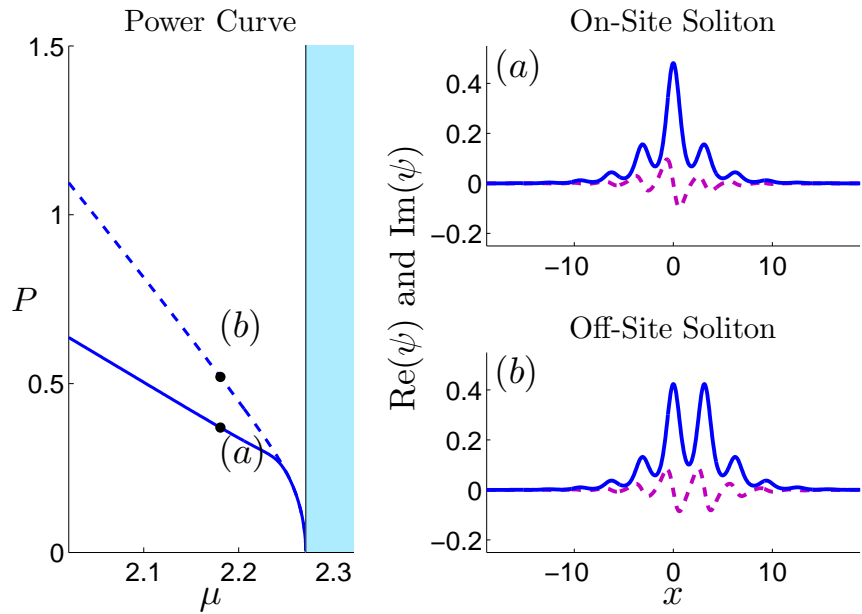}
\caption{(Left) Power curve for the two families of solitons which bifurcate from the lowest band edge of the periodic potential (\ref{Eq:ExLatt})
with $(V_0, W_0)$ given in (\ref{e:V0W0}) and $\sigma=1$.
(Right) Example solitons for the on-site (top) and off-site (bottom) branches at the marked points of the power curve.
Solid lines: Re($\psi$); dashed lines: Im($\psi$).}
 \label{Fig: PowerCurve}
\end{center}
\end{figure}

The above perturbation solution can be continued to all powers in
$\epsilon$, which would seem to suggest that for every $x_0 =
X_0/\epsilon$ there exists a family of solitons parameterized by
$\mu$ that bifurcates from the band edge. This, however, is not the
case; a discrepancy studied in
\cite{Hwang2011,Hwang2012,Pelinovsky2004} for the case of
conservative periodic potentials. Instead, only two possible
solution families exist. For $\mathcal{PT}$-symmetric periodic
potentials they correspond to $x_0 = 0, ~\pi/2$, as shown in
Fig.~\ref{Fig: PowerCurve}. Here the example periodic potential is
(\ref{Eq:ExLatt}) with
\begin{equation} \label{e:V0W0}
V_0=6, \quad W_0=0.25,
\end{equation}
the nonlinearity is chosen to be self-focusing ($\sigma=1$), and
soliton families are represented by their power curves, where the
power is defined as $P(\mu) = \int_{-\infty}^{\infty} |\psi(x;
\mu)|^2 \rd x$. The lower power curve corresponds to solitons
centered at $x_0 = 0$ (referred to as on-site solitons), and the
upper curve corresponds to solitons centered at $x_0 = \pi/2$
(referred to as off-site solitons). On-site solitons are
\PT-symmetric with respect to the origin, while off-site solitons
are \PT-symmetric with respect to a space shift of half-period
$x=\pi/2$. Note that any \PT-symmetric periodic potential is also
\PT-symmetric after a half-period space shift. Hence both on-site
and off-site solitons can be said to be \PT-symmetric in the
underlying \PT-symmetric periodic potential.

The explanation for this apparent contradiction rests in terms that
are exponentially small in $\epsilon$, which cannot be captured by
the above power-series expansion. An exponential-asymptotics
approach was developed for conservatives periodic potentials in
\cite{Hwang2011,Hwang2012,Nixon2013}, based on the method used to
study soliton solutions of the fifth-order Korteweg-de Vries
equation \cite{Akylas1997}. In this article we develop this
exponential asymptotics analysis for complex
$\mathcal{PT}$-symmetric periodic potentials.

\section{The Fourier transform}

The wavepacket solutions of Eq. (\ref{Eq:psi}) in the bandgap are
such that if they decay exponentially in one direction, say upstream
or as $x\rightarrow -\infty$, then they would generically grow
expontentially in the other direction, say downstream or $x
\rightarrow +\infty$. These growing tails are exponentially small in
$\epsilon$. Once we have worked out these exponentially small tail
terms, the center position $x_0$ of the envelope can be determined
by requiring that these terms vanish. To do so, we move to the
Fourier domain, where the exponentially small tail contributions map
to poles with exponentially small strength.

We consider the Fourier transform of the solution $\psi(x, X)$ with
respect to the slow space variable $X$, written formally as
\begin{equation*}
\widehat{\psi}(x,K) = \frac{1}{2\pi} \disp\int_{-\infty}^{\infty} \psi(x,X) \re^{-\ri K X} \rd X.
\end{equation*}
Since $\psi(x,X)$ is complex, we also introduce the Fourier
transform of $\psi^*(x,X)$,
\begin{equation*}
\widehat{\psi^*}(x,K) = \frac{1}{2\pi} \disp\int_{-\infty}^{\infty} \psi^*(x,X) \re^{-\ri K X} \rd X.
\end{equation*}
Substituting the perturbation series (\ref{e:PsixX}) into these
transforms, we get
\begin{equation} \label{Eq:psihat}
\widehat{\psi} = \epsilon  \frac{\alpha \beta}{2} \re^{-\ri K X_0} \sech\left(\frac{\pi}{2} \beta K\right)\left[ p(x) + \ri \epsilon K \nu(x) + \ldots \right],
\end{equation}
and
\begin{equation} \label{Eq:psi*hat}
\widehat{\psi^*} = \epsilon  \frac{\alpha \beta}{2} \re^{-\ri K X_0} \sech\left(\frac{\pi}{2} \beta K\right)\left[ p^*(x) + \ri \epsilon K \nu^*(x) + \ldots \right],
\end{equation}
which are disordered when $\kappa\equiv \epsilon K=O(1)$. Thus we replace them by
uniformly valid expressions
\begin{equation} \label{Eq:Psihat}
\widehat{\psi}(x,\kappa) = \epsilon \re^{-\ri \kappa x_0} \sech\left(\frac{\pi \beta \kappa}{2\epsilon} \right)U_1(x,\kappa; \epsilon),
\end{equation}
and
\begin{equation} \label{Eq:Psi*hat}
\widehat{\psi^*}(x,\kappa) = \epsilon \re^{-\ri \kappa x_0} \sech\left(\frac{\pi \beta \kappa}{2\epsilon} \right)U_2(x,\kappa; \epsilon),
\end{equation}
where
\begin{align*}
U_1(x,\kappa; \epsilon) & =p(x) + \ri \kappa \nu(x) + \ldots, \qquad  |\kappa| \ll 1,  \\
U_2(x,\kappa; \epsilon) & =p^*(x) + \ri \kappa \nu^*(x) + \ldots, \quad |\kappa| \ll 1.
\end{align*}

We now derive the governing equations for $(U_1, U_2)$ by taking the
Fourier transform of equation \eqref{Eq:MultiScale}, which yields
\begin{equation*}
L_0 \widehat{\psi} + 2\ri \kappa \widehat{\psi}_x  - \kappa^2 \widehat{\psi} + \sigma \widehat{\psi^2\psi^*} + \epsilon^2 \eta  \widehat{\psi} = 0.
\end{equation*}
A similar equation can be obtained for $\widehat{\psi^*}$.
Substituting in the expressions \eqref{Eq:Psihat}-\eqref{Eq:Psi*hat}
we find that $(U_1, U_2)$ satisfy
\begin{align}
0=& L_0 U_1 + 2\ri \kappa U_{1x} + (\epsilon^2\eta -  \kappa^2) U_1
+\sigma \: \cosh \frac{\pi \beta \kappa}{2\epsilon} \times  \nonumber \\
& \int_{-\infty}^{\infty} \int_{-\infty}^{\infty} \sech \frac{\pi\beta (\kappa-r)}{2\epsilon}
\sech \frac{\pi\beta (r-s)}{2\epsilon}\sech \frac{\pi\beta s}{2\epsilon} \; U_1(\kappa- r) U_1(r - s) U_2(s)  \: \rd r \rd s,
\label{Eq:U1Full}
\end{align}
\begin{align} 0=& L_0^* U_2 + 2\ri \kappa U_{2x} + (\epsilon^2\eta -
\kappa^2) U_2
+\sigma \: \cosh \frac{\pi \beta \kappa}{2\epsilon} \times  \nonumber \\
& \int_{-\infty}^{\infty} \int_{-\infty}^{\infty} \sech
\frac{\pi\beta (\kappa-r)}{2\epsilon} \sech \frac{\pi\beta
(r-s)}{2\epsilon}\sech \frac{\pi\beta s}{2\epsilon} \; U_2(\kappa-
r) U_2(r - s) U_1(s)  \: \rd r \rd s. \label{Eq:U2Full}
\end{align}

\section{Poles of the Fourier solution}

We are concerned with the behavior near the singularities of
$U_1(x,\kappa ; \epsilon)$ which account for the growing tails of
exponentially small amplitude in the physical space. Singularities
of $U_1$ are expected to occur near values of $\kappa=\kappa_0$
where the linear part of equation \eqref{Eq:U1Full} is zero,
\begin{equation*}
L_0 \phi + 2\ri \kappa_0  \phi_x -  \kappa_0^2 \phi = 0.
\end{equation*}
A change of variables $\phi= \re^{-\ri \kappa_0 x} \widetilde{\phi}$
leaves us with
\begin{equation*}
L_0 \widetilde{\phi} = 0,
\end{equation*}
which has a single bounded solution $\widetilde{\phi} = p(x)$, the
Bloch mode at band edge. Since $\phi$ should have period matching
$p(x)$, $\kappa_{0}$ must be an even integer. This results in poles
when $\kappa_0 = \pm2,\pm4,\pm6, \cdots$. The dominant contributions
to the solution come from the nearest poles to zero at $\kappa_0 =
\pm2$.

Looking for the behavior of the solution $U_1(x,\kappa; \epsilon)$
near $\kappa_0 = 2$, we introduce a local variable
\begin{equation} \label{Eq:xi}
\xi = \frac{\kappa - \kappa_0}{\epsilon},
\end{equation}
that is $\kappa =\kappa_0 + \epsilon \xi$. In this region we expand
the solution to integral equations
\eqref{Eq:U1Full}-\eqref{Eq:U2Full} as (see \cite{Hwang2011})
\begin{equation} \label{Eq:UPoleExpansion}
U_1 = \frac{\re^{-\ri \kappa_0 x}}{\epsilon^4}\left[ \Phi_1(\xi) p(x) +  \epsilon \ri \xi \Phi_1(\xi) \nu(x) + O(\epsilon^2)\right],
\end{equation}
\begin{equation} \label{Eq:U*PoleExpansion}
U_2 = \frac{\re^{-\ri \kappa_0 x}}{\epsilon^4}\left[ \Phi_2(\xi) p^*(x) +  \epsilon \ri \xi \Phi_2(\xi) \nu^*(x) + O(\epsilon^2)\right].
\end{equation}
The dominant contribution to the double integrals in equations
\eqref{Eq:U1Full}-\eqref{Eq:U2Full} comes from three regions: (i) $r
\sim 0$, $s\sim 0$, (ii) $r \sim \kappa$, $s\sim 0$, and (iii) $r
\sim \kappa$, $s\sim \kappa$. Calculation of these contributions
follows that in \cite{Hwang2011}. Under the notations
\begin{equation*}
U_1 =\re^{-\ri \kappa_0 x} {\cal U}_1, \quad U_2 =\re^{-\ri \kappa_0 x} {\cal U}_2,
\end{equation*}
the simplified integral equations in this region become
\begin{align*}
0&=L_0 {\cal U}_1 +  \epsilon 2\ri\xi  {\cal U}_{1x} + \epsilon^2 \left(\eta - \xi^2 \right) {\cal U}_1   \nonumber \\
&+ \frac{\sigma(\alpha\beta)^2}{2 \epsilon^{2}} |p(x)|^2p(x)\int_{-\infty}^{\infty}  \re^{\pi \beta r/2} r \csch\left(\frac{\pi \beta r}{2} \right)
\left[2\Phi_1(\xi - r)+\Phi_2(\xi - r)\right] \rd r,
\end{align*}
\begin{align*}
0&=L_0^* {\cal U}_2 +  \epsilon 2\ri\xi  {\cal U}_{2x} + \epsilon^2 \left(\eta - \xi^2 \right) {\cal U}_2   \nonumber \\
&+ \frac{\sigma(\alpha\beta)^2}{2 \epsilon^{2}} |p(x)|^2p^*(x)\int_{-\infty}^{\infty}  \re^{\pi \beta r/2} r \csch\left(\frac{\pi \beta r}{2} \right)
\left[2\Phi_2(\xi - r)+\Phi_1(\xi - r)\right] \rd r.
\end{align*}
Substituting the expansions
(\ref{Eq:UPoleExpansion})-(\ref{Eq:U*PoleExpansion}) into these
equations, the terms at orders $\epsilon^{-4}$ and $\epsilon^{-3}$
automatically balance. At order $\epsilon^{-2}$, the solvability
conditions yield the following integral equations for $\Phi_1$ and
$\Phi_2$,
\begin{align}   \label{e:Phi1}
(1+ \beta^2\xi^2)\Phi_1 -  \beta^2 \int_{-\infty}^{\infty} \re^{\pi \beta r/2} r \csch\frac{\pi \beta r}{2}
\left[2\Phi_1(\xi - r)+\Phi_2(\xi - r)\right] \rd r =0,
\end{align}
\begin{align}   \label{e:Phi2}
(1+ \beta^2\xi^2)\Phi_2 -  \beta^2 \int_{-\infty}^{\infty} \re^{\pi \beta r/2} r \csch\frac{\pi \beta r}{2}
\left[2\Phi_2(\xi - r)+\Phi_1(\xi - r)\right] \rd r =0.
\end{align}
This is a system of two linear homogeneous Fredholm integral
equations. To solve it, we introduce the integral transform
\begin{equation} \label{e:integraltransf}
\Phi_n(\xi)= \int_0^{\pm \ri \infty} e^{-s \beta \xi}
\phi_n(s) \, \rd s,  \quad n=1, 2,
\end{equation}
where the plus sign is for $\mbox{Im}(\xi)< 0$ and the minus sign
for $\mbox{Im}(\xi)> 0$. Inserting this integral transform into
(\ref{e:Phi1})-(\ref{e:Phi2}) and performing integration by parts,
we find that $(\phi_1, \phi_2)$ satisfy the following differential
equations \cite{Hwang2011}
\begin{equation} \label{e:transformODE1}
\frac{\rd^2 \phi_1}{\rd s^2}+\phi_1 -\frac{4}{\sin^2 \hspace{-0.07cm}
s}\phi_1 -\frac{2}{\sin^2 \hspace{-0.07cm}
s}\phi_2 =0,
\end{equation}
\begin{equation} \label{e:transformODE2}
\frac{\rd^2 \phi_2}{\rd s^2}+\phi_2 -\frac{4}{\sin^2 \hspace{-0.07cm}
s}\phi_2 -\frac{2}{\sin^2 \hspace{-0.07cm}
s}\phi_1 =0.
\end{equation}
Performing dominant balance of these equations around $s\sim 0$ and
requiring the integrals in (\ref{e:integraltransf}) to converge, we
find that $\phi_1$ and $\phi_2$ have the same small-$s$ behavior,
\begin{equation*}
\phi_n \to \chi s^3,  \quad s \to 0 \quad (n=1, 2),
\end{equation*}
where $\chi$ is a certain constant. Together with the fact that
equations (\ref{e:transformODE1})-(\ref{e:transformODE2}) are
symmetric in $\phi_1$ and $\phi_2$, we see that solutions $\phi_1$
and $\phi_2$ must be identical, thus
\begin{equation*}
\Phi_1(\xi)=\Phi_2(\xi)\equiv \Phi(\xi).
\end{equation*}
Under this reduction, integral equations
(\ref{e:Phi1})-(\ref{e:Phi2}) reduce to
\begin{align}
(1+ \beta^2\xi^2)\Phi - 3\beta^2 \int_{-\infty}^{\infty} \re^{\pi \beta r/2} r \csch\frac{\pi \beta r}{2}
\Phi(\xi - r) \rd r =0,
\end{align}
which has been solved before \cite{Hwang2011,Akylas1997}. Its exact
solution in the region $\left|\Imag(\xi)\right| \geq 1/\beta$ is
given by
\begin{equation} \label{Eq:Phi0solution}
\Phi(\xi) = \frac{6\beta^4}{1+ \beta^2 \xi^2} \int_0^{\pm \ri
\infty} \frac{1}{\sin^2 s}\phi(s) \re^{-s\beta \xi} \rd s,
\end{equation}
\begin{equation*}
\phi(s) = C\left( \frac{2}{\sin s} + \frac{\cos^2 s}{\sin s} -
\frac{3s \cos s}{\sin^2s}\right),
\end{equation*}
where $C$ is a constant. Clearly this solution has simple-pole
singularities at $\xi = \pm \ri /\beta$. Since the integral of
(\ref{Eq:Phi0solution}) at these points is equal to $-C/6$, we see
that
\begin{equation*}
\Phi(\xi) \sim -\frac{C\beta^4}{1+ \beta^2 \xi^2},~~{\rm for} ~\xi
\to  \pm \frac{\ri}{\beta}.
\end{equation*}
Utilizing this equation and recalling the scaling (\ref{Eq:xi}), we
obtain the following singular behaviors for the Fourier solutions
$\widehat{\psi}(x, K)$ and $\widehat{\psi^*}(x, K)$,
\begin{equation} \label{e:eqsingularities1}
\widehat\psi(x, K) \sim \frac{\beta^3 C}{\epsilon^3}
e^{-\pi\beta\kappa_0/2\epsilon}e^{\mp
X_0/\beta} \frac{e^{-i\kappa_0(x+x_0)}}{K-\frac{\kappa_0}{\epsilon} \pm
\frac{i}{\beta}}p(x) \qquad (K\to  \frac{\kappa_0}{\epsilon} \mp
\frac{i}{\beta}),
\end{equation}
\begin{equation}
\widehat{\psi^*}(x, K) \sim \frac{\beta^3 C}{\epsilon^3}
e^{-\pi\beta\kappa_0/2\epsilon}e^{\mp
X_0/\beta}\frac{e^{-i\kappa_0(x+x_0)}}{K-\frac{\kappa_0}{\epsilon} \pm
\frac{i}{\beta}}p^*(x) \qquad (K\to  \frac{\kappa_0}{\epsilon} \mp
\frac{i}{\beta}).
\end{equation}
Then from the symmetry relation $[\widehat\psi(x,
K)]^*={\widehat\psi}^*(x, -K^*)$ of complex functions $\psi(x, X)$,
it follows that
\begin{equation} \label{e:eqsingularities2}
\widehat\psi(x, K) \sim -\frac{\beta^3 C^*}{\epsilon^3}
e^{-\pi\beta\kappa_0/2\epsilon}e^{\mp X_0/\beta}
\frac{e^{i\kappa_0(x+x_0)}}{K+\frac{\kappa_0}{\epsilon} \pm
\frac{i}{\beta}} p(x)  \qquad  (K\to  -\frac{\kappa_0}{\epsilon} \mp
\frac{i}{\beta}).
\end{equation}

It remains to determine the constant $C$. To do so, we match the
large-$\xi$ asymptotics of $U_1$ to the solution $U_1$ in the region
where $\kappa = O(1)$ but away the singularity at $\kappa \approx
\kappa_0$. Notice that for $|\xi| \rightarrow \infty$ the main
contribution of the integral in solution \eqref{Eq:Phi0solution}
comes from the region $s\sim 0$ where $\phi(s) \sim \frac{2}{5} C
s^3$. This yields
\begin{equation*}
\Phi(\xi) \rightarrow \frac{12 C}{5} \frac{1}{\xi^4},~~{\rm for} ~ |\xi| \rightarrow \infty.
\end{equation*}
Putting this in terms of $U_1$ and $U_2$, we find the follow
asymptotic behaviors for the inner solutions around $\kappa \sim
\kappa_0$,
\begin{equation} \label{Eq:InnerAsymptotics}
U_1(x,\kappa) \sim \frac{12 C}{5} \frac{p(x)}{(\kappa - \kappa_0)^4}\re^{-\ri \kappa_0 x}, \qquad \kappa \sim \kappa_0,
\end{equation}
and
\begin{equation} \label{Eq:InnerAsymptotics2}
U_2(x,\kappa) \sim \frac{12 C}{5} \frac{p^*(x)}{(\kappa - \kappa_0)^4}\re^{-\ri \kappa_0 x}, \qquad \kappa \sim \kappa_0.
\end{equation}
This behavior we match to the solution of
\eqref{Eq:U1Full}-\eqref{Eq:U2Full} in the outer region, $\kappa =
O(1)$, in section \ref{Sec: Outer}.

\section{Fourier solution away from poles}
\label{Sec: Outer}

Since the strength of the poles cannot be determined uniquely from
the local behavior we look to match the asymptotic behavior of the
near pole solutions (\ref{Eq:InnerAsymptotics}) to the solution of
equations \eqref{Eq:U1Full}-\eqref{Eq:U2Full} for $\kappa = O(1)$
away from these poles. The main contribution of the double integrals
in \eqref{Eq:U1Full}-\eqref{Eq:U2Full} now comes from the triangular
region $0< s <\kappa$, $0< s < r$ when $\kappa >0$ and $\kappa<s<0$,
$r<s<0$ when $\kappa < 0$. Over this region,
\begin{equation*}
\cosh \frac{\pi \beta \kappa}{2\epsilon} \: \sech \frac{\pi\beta (\kappa-r)}{2\epsilon} \:
\sech \frac{\pi\beta (r-s)}{2\epsilon}\: \sech \frac{\pi\beta s}{2\epsilon} \approx 4,
\end{equation*}
thus Eqs. \eqref{Eq:U1Full}-\eqref{Eq:U2Full}, to leading order of
$\epsilon$, reduce to
\begin{align}
0&=L_0 U_1 + 2\ri\kappa U_{1x} - \kappa^2 U_1 + 4\sigma \int_0^\kappa \int_0^r U_1(x,\kappa-r)U_1(x,r-s) U_2(x,s)\rd s \rd r,   \label{e:Outer1} \\
0&=L_0^* U_2 + 2\ri\kappa U_{2x} - \kappa^2 U_2 + 4\sigma \int_0^\kappa \int_0^r U_2(x,\kappa-r)U_2(x,r-s) U_1(x,s)\rd s \rd r.  \label{e:Outer2}
\end{align}
We propose to solve these outer integral equations numerically
\cite{Nixon2013}. Since this is a Volterra integral equation, it can
be easily tackled by explicit numerical methods. First we discretize
$\kappa$, $\kappa_n = n \Delta \kappa$, and write
\begin{align*}
U_1(x, \kappa_n) &= U^{(n)}_1(x), \\
U_2(x, \kappa_n) &= U^{(n)}_2(x).
\end{align*}
Then we approximate the double convolution using the trapezoidal
rule. After the terms in the resulting equations are rearranged,
$(U^{(n)}_1, U^{(n)}_2)$ are found to satisfy the following linear
inhomogeneous equations
\begin{align}
\left[L_0  + 2\ri\kappa_n \partial_x - \kappa_n^2 + 2 \sigma U^{(0)}_1U^{(0)}_2 (\Delta
\kappa)^2 \right] U^{(n)}_1&= - 4\sigma F^{(n)}_1,   \label{e:Un1}  \\
\left[L^*_0  + 2\ri\kappa_n \partial_x - \kappa_n^2 + 2\sigma U^{(0)}_1U^{(0)}_2 (\Delta
\kappa)^2 \right] U^{(n)}_2& = - 4\sigma F^{(n)}_2,   \label{e:Un2}
\end{align}
where the inhomogeneous terms $F^{(n)}_1$ and $F^{(n)}_2$ are given
by
\begin{align*}
F^{(n)}_1 &= \Delta \kappa^2 \left[ \frac{1}{2} U^{(0)}_2 I^{(n)}_1 + \displaystyle
\sum_{m=1}^{n-1} U^{(m)}_2 I^{(n-m)}_{1}\right], \\
F^{(n)}_2 &= \Delta \kappa^2 \left[ \frac{1}{2} U^{(0)}_1 I^{(n)}_2 + \displaystyle
\sum_{m=1}^{n-1} U^{(m)}_1 I^{(n-m)}_{2}\right], \\
I^{(m)}_j &=\displaystyle\sum_{l = 0}^{m-1}U^{(l)}_j U^{(m-l)}_j, \quad {\rm for}~ 1\le m<n, \hspace{0.3cm} j=1, 2, \\
I^{(n)}_j &= \displaystyle\sum_{l = 1}^{n-1}U^{(l)}_j U^{(n-l)}_j, \quad j=1, 2.
\end{align*}
The initial conditions $U^{(0,1)}_1$ and $U^{(0,1)}_2$ can be
obtained from equations \eqref{Eq:psihat}-\eqref{Eq:Psi*hat} as
\begin{align*}
U^{(0)}_1(x) &=\frac{\alpha \beta}{2} p(x), \\
U^{(1)}_1(x)& =\frac{\alpha \beta}{2}\left[p(x) + \ri \Delta \kappa~
\nu(x)\right], \\
U^{(0)}_2(x) &=\frac{\alpha \beta}{2} p^*(x), \\
U^{(1)}_2(x)& =\frac{\alpha \beta}{2}\left[p^*(x) + \ri \Delta \kappa~
\nu^*(x)\right].
\end{align*}

Unlike the case of conservative periodic potentials
\cite{Hwang2011}, the linear operators on the left sides of
equations (\ref{e:Un1})-(\ref{e:Un2}) are non-Hermitian, thus
conjugate gradient iterations for solving them will not work here.
Instead we apply the adjoint linear operators to these equations and
turn them into normal equations, which can then be solved by
preconditioned conjugate gradient iterations.

Recalling the asymptotic behaviors of inner solutions in equations
(\ref{Eq:InnerAsymptotics})-(\ref{Eq:InnerAsymptotics2}), we are
motivated to introduce the change of variables
\begin{equation*}
\widetilde{U}_1(x, \kappa) = (\kappa-\kappa_0)^{4}
U_1(x,\kappa),  \quad \widetilde{U}_2(x, \kappa) = (\kappa-\kappa_0)^{4}
U_2(x,\kappa).
\end{equation*}
Then
\begin{equation} \label{Eq:UtildeAsymptotics}
\widetilde{U}_1(x,\kappa) \to \frac{12 C}{5} p(x) \re^{-\ri
\kappa_0 x},  \quad ~ \mbox{as} \ \kappa \to \kappa_0,
\end{equation}
\begin{equation} \label{Eq:Utilde*Asymptotics}
\widetilde{U}_2(x,\kappa) \to \frac{12 C}{5} p^*(x) \re^{-\ri
\kappa_0 x},  \quad \mbox{as} \ \kappa \to \kappa_0.
\end{equation}

At this point we note that $p(x)$ and $\nu(x)$ are both
\PT-symmetric due to the \PT-symmetric periodic potential $V(x)$.
Thus the initial conditions for $U_1$ and $U_2$ are both
$\mathcal{PT}$-symmetric. So are the outer equations
(\ref{e:Outer1})-(\ref{e:Outer2}) as well. Thus, we see from
equations (\ref{Eq:UtildeAsymptotics})-(\ref{Eq:Utilde*Asymptotics})
that $C$ must be a real constant.

\begin{figure}[htbp!]
\begin{center}
\includegraphics[width=4.5in]{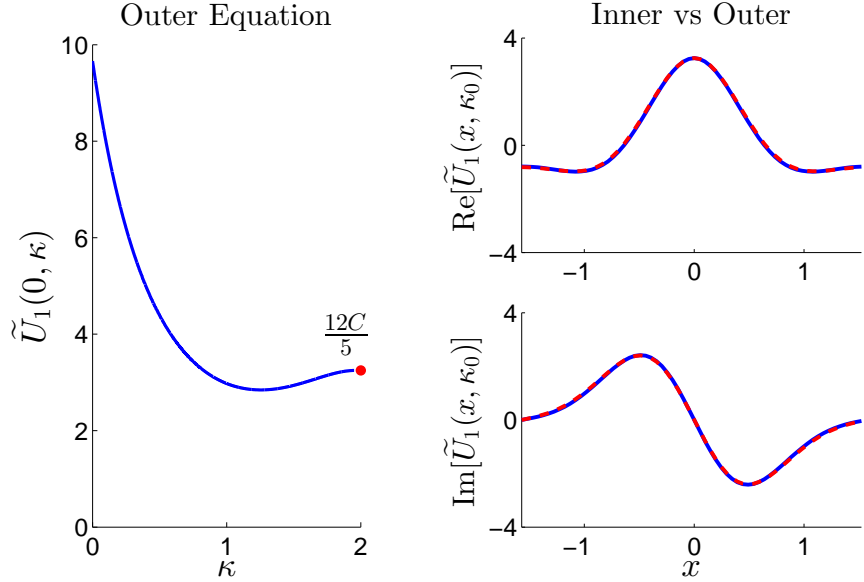}
\caption{(Left) Numerical solution $\widetilde{U}_1(x, \kappa)$ versus $\kappa$ at $x=0$, which converges to the value $\frac{12C}{5}$
when $\kappa\to \kappa_0=2$. (Right) Comparison of the numerical outer-solution profile $\widetilde{U}_1(x, \kappa_0)$ (dashed red)
with the analytical prediction (\ref{Eq:UtildeAsymptotics}) (solid blue). The
periodic potential used is (\ref{Eq:ExLatt}) with $(V_0, W_0)$ given in (\ref{e:V0W0}), and $\sigma=1$.}
\label{Fig: OuterEquation}
\end{center}
\end{figure}

In figure \ref{Fig: OuterEquation} we verify all this numerically.
The periodic potential used here is (\ref{Eq:ExLatt}) with $(V_0,
W_0)$ values given in (\ref{e:V0W0}), and $\sigma=1$ (self-focusing
nonlinearity). For this potential,
\begin{equation*}
\alpha \approx 1.6830, \quad \beta \approx 0.7177.
\end{equation*}
On the left, we plot the numerical solution $\widetilde{U}_1(x,
\kappa)$ versus $\kappa$ at $x=0$. As predicted in
(\ref{Eq:UtildeAsymptotics}), this solution indeed approaches a
finite theoretical value $\frac{12C}{5}$ when $\kappa\to \kappa_0=2$
(recall that $p(x=0)$ has been scaled to one). This numerical curve
allows us to determine the $C$ value as
\begin{equation}
C \approx 1.35.   \label{Eq:CValue}
\end{equation}
In section \ref{Sec: Eigenvalues} we are able to independently
verify this value of $C$ quantitatively, by comparing our prediction
to linear-stability eigenvalues with numerically computed
eigenvalues. On the right side of figure 2, we plot the profile of
the numerical outer-solution profile $\widetilde{U}_1(x, \kappa_0)$
(dashed red) with the analytical prediction
(\ref{Eq:UtildeAsymptotics}) (solid blue). The dashed and solid
curves are almost indistinguishable, confirming the agreement
between numerical solutions and analytical predictions.

\section{The inverse Fourier transform}

The physical solution $\psi(x, X)$ is obtained by taking the inverse
Fourier transform of $\widehat\psi(x, K)$,
\begin{equation*}
\psi(x,X)=\int_{\cal C} \widehat\psi(x,K) e^{iKX}\rd K\,.
\end{equation*}
If we require this physical solution to decay upstream ($x\to
-\infty$), then the contour $\cal C$ in this inverse Fourier
transform should be taken along the line $\Imag(K)=-1/\beta$ and
pass below the poles $K= \pm \kappa_0/\epsilon - i/\beta$. It should
also pass above the pole $K=-i/\beta$ of the $\sech(\pi\beta K/2)$
term in Eq. (\ref{Eq:Psihat}). Then when $x\gg 1$ (downstream), by
completing the contour $\cal C$ with a large semicircle in the upper
half plane, we pick up dominant contributions from the pole
singularities at $K=\pm\kappa_0/\epsilon -i/\beta$ and $K=-i/\beta$.
Utilizing the pole-singularity solutions (\ref{e:eqsingularities1}),
(\ref{e:eqsingularities2}), collecting these pole contributions and
recalling the reality of $C$, the wave profile of the solution far
downstream is then found to be
\begin{align}
\psi\sim 2\epsilon \alpha \hspace{0.05cm} e^{-(X-X_0)/\beta} p (x)
+\frac{4\pi\beta^3 C}{\epsilon^3}
e^{- \pi \beta /\epsilon}\sin(2 x_0)
e^{(X-X_0)/\beta}  p(x), \quad \,\, x \gg 1/\epsilon. \label{Eq:Downstream}
\end{align}

For this solution to be a solitary wave, the growing term in
(\ref{Eq:Downstream}) must vanish so $\sin \left(2 x_0 \right) = 0$.
Thus, we find that there are two allowable locations for solitons
(relative to the periodic potential),
\begin{equation*}
x_0 = 0, \quad \pi/2.
\end{equation*}
That is, the envelope of the solitons must be located either at the
point of \PT-symmetry ($x_0=0$), or half-a-period away from the
point of \PT-symmetry ($x_0=\pi/2$). The resulting two families of
solitons are also \PT-symmetric with respect to either $x_0=0$ or
$x_0=\pi/2$. These two soliton families are precisely the ones which
were found numerically for the sample periodic potential
(\ref{Eq:ExLatt}) in figure 1.

Finally, if one wishes to obtain wave packets $\psi(x,X)$ which
decay for $x\to +\infty$ but contain a growing tail for $x\ll -1$,
then the contour $\cal C$ in the inverse Fourier transform should be
taken along the line $\Imag(K)=1/\beta$ and pass above the poles $K=
\pm \kappa_0/\epsilon + i/\beta$ and below the pole $K=i/\beta$.
Then when $x\ll -1$, by completing the contour $\cal C$ with a large
semicircle in the lower half plane and picking up dominant pole
contributions from equations (\ref{e:eqsingularities1}) and
(\ref{e:eqsingularities2}), the wave profile of the solution for
$x\ll -1$ is found to be
\begin{align}
\psi\sim  2\epsilon \hspace{0.04cm} \alpha \hspace{0.05cm}
e^{(X-X_0)/\beta} p (x) -\frac{4\pi\beta^3 C}{\epsilon^3} e^{- \pi \beta /
\epsilon}\sin(2 x_0) e^{-(X-X_0)/\beta} p (x),
\quad \,\, x \ll -1/\epsilon. \label{Eq:Downstream2}
\end{align}

\section{Connection to stability}
\label{Sec: Eigenvalues}

We now turn our attention to the linear-stability problem of these
two soliton families near band edges. When these solitons bifurcate
out from a band edge, a pair of exponentially small eigenvalues
bifurcate out from the origin \cite{Hwang2011}. Using the tail
formula \eqref{Eq:Downstream} we are able derive an analytic
approximation for these eigenvalues, and thus, by comparison with
the numerically computed eigenvalues, verify the validity of our
formula for the exponentially small tail as well as the numerical
computation of the outer equation.

Let $\psi_s(x)= \psi(x; x_{0s})$ be a soliton solution of equation
\eqref{Eq:psi} with center at $x_0 = x_{0s}$, which decays to zero
as $x \rightarrow \pm \infty$. To study its linear stability we
perturb this soliton as
\begin{equation*}
\Psi = \re^{-\ri \mu t} \left[ \psi_s + w_1(x) \hspace{0.05cm}
\re^{ \lambda t} + w_2^*(x) \hspace{0.05cm} \re^{ \lambda^* t}
\right],
\end{equation*}
where $|w_1|,  |w_2|\ll |\psi_s|$. After substitution into equation
\eqref{Eq:psi} and linearizing, we arrive at the eigenvalue problem
\begin{equation} \label{Eq:EigenProblem}
\ri {\cal L} W = \lambda  W,
\end{equation}
where $W = [w_1 ~ w_2]^T$ (the superscript `$T$' representing
transpose of a vector),
\begin{align*}
{\cal L} &= \left( \begin{array}{c c} L_{11} &  L_{12} \\
L_{21} & L_{22}\end{array} \right), \\
L_{11} &= \partial_{xx} +\mu   -  V(x)  + 2 \sigma |u|^2,  \\
L_{12} &= \sigma u^2, \\
L_{21} &= -\sigma \left(u^2\right)^*,  \\
L_{22} &= -\left(\partial_{xx} +\mu  -  V^*(x)  + 2\sigma |u|^2
\right).
\end{align*}
Unlike the eigenvalue problem for conservative potentials
\cite{Hwang2011}, here the eigenvalue problem cannot be reduced to
one with zero diagonal elements. Thus our treatment will require
that we solve a coupled equation at each step.

As we've seen, the soliton near the band edge is a low-amplitude
wave packet whose envelope is governed by \eqref{Eq:Envelope}. While
the envelope is translationally invariant, this symmetry is not
shared by the full equation \eqref{Eq:NLS}. As such, the zero
eigenvalue associated with this translation invariance bifurcates
due to the broken symmetry. The eigenvalue we are looking for is
exponentially small as $\epsilon \rightarrow 0$, and we construct
$W$ as a series expansion in $\lambda$,
\begin{equation*}
W =  W_0 + \lambda W_1 + \lambda^2 W_2 + \cdots.
\end{equation*}
In this case, the equations at the first few orders of $\lambda$ are
\begin{equation*}
\ri {\cal L} W_0 = 0,~~~~\ri {\cal L} W_1 = W_0,~~~~\ri {\cal L} W_2 = W_1.
\end{equation*}

At $O(1)$ the solution for $W_0$ is
\begin{equation*}
W_0 = \left[\left. \frac{\partial \psi}{\partial x_0}\right|_{x_0=x_{0s}} ~~~ \left.\left(\frac{\partial \psi}{\partial x_0}\right)^*\right|_{x_0=x_{0s}}
\right]^T,
\end{equation*}
which can be verified by taking the derivative of \eqref{Eq:psi}
with respect to the center parameter $x_0$. To first order in
$\epsilon$ the $W_0$ solution may be approximated as
\begin{equation*}
\left. \frac{\partial \psi}{\partial x_0}\right|_{x_0=x_{0s}} \sim - \epsilon^2 A'(X) p(x).
\end{equation*}
If we use the asymptotic formula for the downstream tail
\eqref{Eq:Downstream} we find that for $x \gg 1/\epsilon$, $W_0$
contains a growing tail
\begin{equation*}
\left. \frac{\partial \psi}{\partial x_0}\right|_{x_0=x_{0s}} \sim  \cos(2 x_{0s})
\frac{8 \pi \beta^3 C}{\epsilon^3} \re^{-\pi \beta/ \epsilon} \re^{(X-X_{0s})/\beta} p(x), \quad x \gg 1/\epsilon.
\end{equation*}

At $O(\lambda)$ we set
\begin{equation*}
W_1 = [ \ri w^{(1)}~~~ -\ri w^{(1)*} ]^T,
\end{equation*}
and further expand $w^{(1)}$ as a perturbation series in $\epsilon$,
\begin{equation*}
w^{(1)} = B(X) p(x) + \epsilon B'(X) \nu(x) + \epsilon^2 \widetilde{w}^{(1)} + \cdots.
\end{equation*}
Inserting this expansion into $\ri {\cal L} W_1 = W_0$ we find that
at order $1$ and $\epsilon$ the equation is automatically satisfied,
and at order $\epsilon^2$ we have
\begin{equation*}
L_0 \widetilde{w}^{(1)} = -[p(x) + 2\nu'(x)] B''(X) - \eta B(X) - \sigma A^2(X) B(X) p^3(x) + A'(X) p(x).
\end{equation*}
The solvability condition for this equation along with the
definition for $a$ and $D$ given in \eqref{Eq:aD} now results in the
following inhomogeneous differential equation for $B$,
\begin{equation*}
D B'' + \eta B + \sigma a A^2 B  = A',
\end{equation*}
whose solution is
\begin{equation*}
B(X) = \frac{1}{2D} (X-X_{0s}) A(X).
\end{equation*}

Proceeding to order $\lambda^2$ we let
\begin{equation*}
W_2 = [  w^{(2)}~~~  w^{(2)*} ]^T,
\end{equation*}
and again expand the solution as a series in $\epsilon$,
\begin{equation*}
w^{(2)} = \frac{1}{\epsilon^2} \left[F(X) p(x) + \epsilon F'(X) \nu(x) + \epsilon^2 \widetilde{w}^{(2)} + \dots\right].
\end{equation*}
Inserting this expansion into $\ri {\cal L} W_2 = W_1$, at order
$\epsilon^2$ we arrive at the equation
\begin{align*}
L_0 \widetilde{w}^{(2)} = & -[p(x) + 2 \nu'(x) ] F''(X) -\eta p(x) F(X) \nonumber \\
& - 3 \sigma A^2(X) F(X) p^3(x) + \frac{1}{2D} (X-X_{0s})A(X) p(x)
\end{align*}
for $\widetilde{w}^{(2)}$, which has solvability condition
\begin{equation} \label{Eq:F}
D F'' + \eta F + 3 \sigma a A^2 F = \frac{1}{2D} (X-X_{0s} ) A.
\end{equation}
If we require that $F$ decay upstream, then this will result in a solution that grows exponentially downstream
\begin{equation*}
F(X) \sim R\: \re^{(X-X_{0s})/\beta}, ~~~X \gg 1
\end{equation*}
for some constant $R$. By multiplying both sides of \eqref{Eq:F} by
the homogenous solution $A'(X)$ and then integrating from $-\infty$
to $X$ we get
\begin{equation} \label{Eq:IntF}
\int_{-\infty}^{X} A'\left[DF'' + \eta F + 3 \sigma a A^2 F \right] \rd \widetilde{X} = \frac{1}{2D} \int_{-\infty}^{X} (X-X_{0s})A A' \rd \widetilde{X}.
\end{equation}
After substituting in the expression for $A$ given in
\eqref{Eq:Asech} as well as the asymptotic behavior of $F$, we
perform integration by part on both sides of equation
\eqref{Eq:IntF} to find $R = \alpha/8\beta$. This, in turn, gives us
\begin{equation*}
w^{(2)} \sim \frac{1}{\epsilon^2} \frac{\alpha}{8 \beta} \re^{(X-X_{0s})/\beta} p(x), \quad X \gg 1.
\end{equation*}

We can now balance the growing tail in $W_0$ with the growing tail
in $W_2$ so that $W$ is bounded. This balancing yields
\begin{equation}  \label{Eq:lambda}
\lambda^2 =  -\cos(2 x_{0s}) C \frac{64 \pi \beta^4 }{\alpha
\epsilon} \re^{-\pi \beta / \epsilon}.
\end{equation}
Thus, this eigenvalue is imaginary (stable) for the soliton family with $\cos(2
x_{0s}) C>0$ and real (unstable) for the other soliton family, and its
magnitude is exponentially small. In this paper, the soliton family
with $\cos(2 x_{0s}) C>0$ is called the on-site family, and the one
with $\cos(2 x_{0s}) C<0$ is called the off-site family.

\begin{figure}[htbp!]
\begin{center}
\includegraphics[width=4.5in]{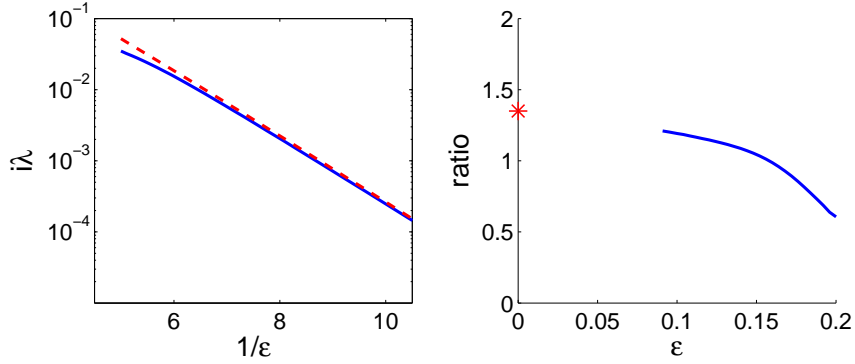}
\caption{(Left) Comparison of numerical (solid blue) and analytical (dashed red) eigenvalues for on-site solitons of figure 1. (Right) Ratio
of $-\lambda^2/[64 \pi \beta^4 \re^{-\pi \beta /\epsilon}/\alpha \epsilon]$ versus $\epsilon$
using numerical $\lambda$ values. The red asterisk marks the $C$ value (\ref{Eq:CValue}) from outer-equation computation.}
\label{Fig: EigCompare}
\end{center}
\end{figure}

A comparison of the analytical prediction (\ref{Eq:lambda}) to
numerical eigenvalues for the family of on-site solitons (with
$x_{0s}=0$) from figure 1 is shown in figure \ref{Fig: EigCompare}
for the example potential \eqref{Eq:ExLatt}. We can see from the
left panel that the numerical eigenvalues are well approximated by
the analytic prediction. Notice that the analytical eigenvalue
formula contains the constant $C$, and the ratio of $-\lambda^2/[64
\pi\beta^4 \re^{-\pi \beta/\epsilon}/\alpha \epsilon]$ approaches
$C$ as $\epsilon\to 0$. In the right panel of figure \ref{Fig:
EigCompare} we find that this ratio indeed approaches the value of
$C$ found in the earlier \eqref{Eq:CValue}. This gives us a
quantitative verification for the asymptotic formula
\eqref{Eq:Downstream} for the downstream behavior of $\psi(x)$.

The above eigenvalue calculations show that the off-site solitons
near band edges are always linearly unstable due to the unstable
eigenvalue from (\ref{Eq:lambda}). For on-site solitons near band
edges, the eigenvalues from (\ref{Eq:lambda}) are stable. However,
other unstable eigenvalues may appear which can still render the
soliton unstable. For instance, if part of the Bloch bands of the
periodic potential is complex, then the linear-stability operator
$\ri {\cal L}$ will possess unstable continuous eigenvalues. For
another instance, if some segment of $\ri {\cal L}$'s continuous
eigenvalues is embedded inside another segment of $\ri {\cal L}$'s
continuous eigenvalues on the imaginary axis when $\epsilon=0$, then
when $\epsilon\ne 0$, complex discrete eigenvalues might also
bifurcate out from the imaginary axis. For on-site solitons
originating from the lowest Bloch-band edge into the semi-infinite
gap, continuum-embedding does not occur. In this case, when the
Bloch bands of the periodic potential are all-real, then the family
of on-site solitons near this band edge are indeed linearly stable.
However, for on-site solitons originating from other band edges,
their stability still needs careful examination.

\section{Construction of multi-soliton bound states}
\label{Sec: Boundstates}

By eliminating the growing tail terms we are able to locate the
center of the soliton envelope, however more information than just
this can be gleaned from the exponentially small tail terms. By
matching the downstream growing tail of a wavepacket with the
upstream decaying tail of another neighboring wavepacket we can
analytically construct multi-soliton bound states. This technique
has been used successfully for the construction of multi-soliton
bound states for one- and two-dimensional conservative systems
\cite{Akylas2012,Hwang2012,Nixon2013,Akylas1997}, and the same basic
analysis holds for the $\mathcal{PT}$ symmetric case.

Consider two wavepackets centered at $x_1= X_1/\epsilon$ (left) and
$x_2= X_2/\epsilon $ (right) respectively. The left wavepacket
decays for $x - x_1 \ll -1$ and has a growing exponential tail for
$x-x_1 \gg 1$, and the right wavepacket decays for $x-x_2 \gg 1$ and
has a growing exponential tail for $x - x_2 \ll -1$. In the
intermediate range, $x_1 \ll x \ll x_2$, the decaying and growing
tails of the two wavepackets must match in order for them to form a
bound state. In this matching region, the left wavepacket's
asymptotics is given by (\ref{Eq:Downstream}) with $x_0$ replaced by
$x_1$, and the right wavepacket's asymptotics is given by
(\ref{Eq:Downstream2}) with $x_0$ replaced by $x_2$. Matching of
these asymptotics results in the following system of equations
\begin{align*}
2\epsilon \alpha \re^{-(X-X_1)/\beta} p(x) &= \mp \frac{4 \pi
\beta^3C }{\epsilon^3}\re^{ - \pi \beta / \epsilon} \sin
\left( 2x_2  \right) \re^{-(X-X_2)/\beta} p(x),
\\
2\epsilon \alpha \re^{(X-X_2)/\beta} p(x)&= \pm \frac{4 \pi
\beta^3C} {\epsilon^3} \re^{ - \pi \beta  / \epsilon}
\sin \left(2 x_1\right) \re^{(X-X_1)/\beta}
p(x),
\end{align*}
where `$\mp$' comes from a possible $\pi$ phase shift between the
neighboring wavepackets. After some reductions, this set of
equations become
\begin{equation}
\sin\left( 2 x_1 \right) = -\sin\left( 2
x_2  \right) = \pm \frac{\alpha\epsilon^4}{2 \pi \beta^3
C } \re^{\pi\beta/\epsilon} \re^{\epsilon
(x_1-x_2)/\beta}, \label{Eq:Centers}
\end{equation}
which are identical to the matching conditions found previously
\cite{Akylas2012}. As was explained there, this system of
equations admits an infinite number of solutions for each fixed
$\epsilon>0$. By varying $\epsilon$, infinite families of
two-soliton bound states are obtained. Furthermore, for equation
\eqref{Eq:Centers} to admit solutions the absolute value of the
right-hand side must be less than or equal to 1. Since this is not
the case in the limit $\epsilon \rightarrow 0$ for any finite
separation, $x_2-x_1$, we see that every solution family bifurcates
at some critical distance away from the band edge, i.e., when
$\epsilon$ is above a certain critical value $\epsilon_c$.

Equation (\ref{Eq:Centers}) admits two types of solutions. The first
type is
\begin{equation}  \label{e:x1x21}
x_1 = \frac{\pi}{2} -d, \quad x_2 = N\pi +\frac{\pi}{2} +d
\end{equation}
for the plus sign (same envelope polarity) and
\begin{equation}  \label{e:x1x22}
x_1 = -d, \quad x_2 = N\pi +d
\end{equation}
for the minus sign (opposite envelope polarity), where $N$ is an
integer, and $0<d<\pi/2$ is determined from the equation
\begin{equation}
\sin 2d=\frac{\alpha\epsilon^4}{2 \pi \beta^3
C } \re^{\pi\beta/\epsilon} \re^{-\epsilon
(N\pi+2d)/\beta}.
\label{e:conditionsine}
\end{equation}
For the envelope positions (\ref{e:x1x21}), the resulting
two-soliton bound state $\psi(x)$ is \PT-symmetric with respect to
the bound-state center $x_c=(N+1)\pi/2$; for the envelope positions
(\ref{e:x1x22}), $i\psi(x)$ is \PT-symmetric with respect to the
bound-state center $x_c=N\pi/2$. In both cases, the bound states are
\PT-symmetric (after a horizonal spatial shift).

Numerically we have confirmed the existence of this type of bound
states. To demonstrate, we take the periodic potential
(\ref{Eq:ExLatt}) with $(V_0, W_0)$ given in (\ref{e:V0W0}), and
$\sigma=1$. Then for $N=5$ and same envelope polarity, this family
of two-soliton bound states in the semi-infinite gap is found and
presented in figure 4. The power curve (upper left panel) has two
connected branches which do not touch the band edge, in agreement
with the analysis. The bound states on the lower power branch (lower
left panel) are roughly given by two on-site solitons, while the
bound states on the upper power branch (lower right panel) are
roughly given by two off-site solitons. Notice that these bound
states are \PT-symmetric with respect to a spatial shift of $3\pi$.
When $N$ is varied, the analytical and numerical critical
$\epsilon_c$ values (in the upper right panel) are also consistent.
In addition, their agreement improves for larger $N$, which is
expected.

\begin{figure}[htbp!]
\begin{center}
\includegraphics[width=5in]{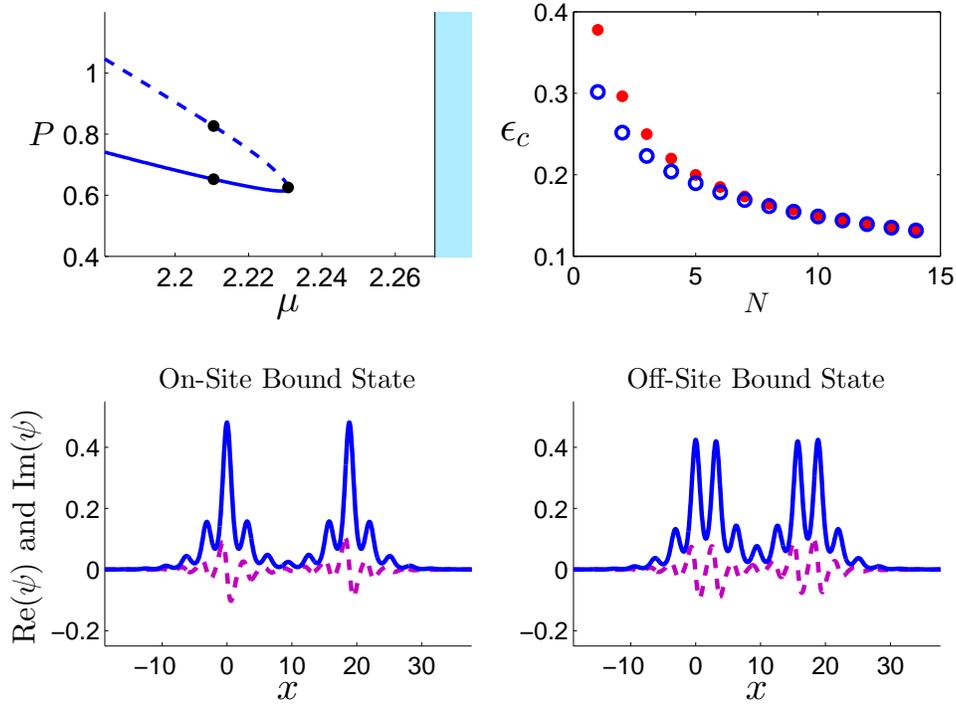}
\end{center}
\caption{(Upper left) Power curve for the solution family with $N=5$ and same envelope polarity in the semi-infinite gap. (Upper right)
Comparison of analytically predicted critical values $\epsilon_c$ (blue circles)
with numerically obtained values (red dots) at various $N$ values. (Lower panels)
Example bound-state solutions at the marked points of the lower and upper power branches. The
periodic potential used is (\ref{Eq:ExLatt}) with $(V_0, W_0)$ given in (\ref{e:V0W0}), and $\sigma=1$.}
\label{Fig: MultiSoliton}
\end{figure}

The second type of solutions admitted by equation \eqref{Eq:Centers}
is
\begin{equation} \label{e:x1x21b}
x_1 = -\frac{\pi}{2} -d, \quad x_2 = N\pi -d
\end{equation}
for the plus sign (same envelope polarity) and
\begin{equation}  \label{e:x1x22b}
x_1 = -d, \quad x_2 = \left(N+\frac{1}{2}\right)\pi -d
\end{equation}
for the minus sign (opposite envelope polarity), where $N$ is an
integer and $0<d<\pi/2$ is determined from the equation
\begin{equation}
\sin 2d=\frac{\alpha\epsilon^4}{2 \pi \beta^3
C } \re^{\pi\beta/\epsilon} \re^{-\epsilon
(N+\frac{1}{2})\pi/\beta}.
\label{e:conditionsine2}
\end{equation}
For these envelope positions, the resulting two-soliton bound states
comprise roughly an on-site and off-site solitons, and are thus
always non-\PT-symmetric under any spatial shift. These
non-\PT-symmetric bound states, if they were to exist, would
contradict the analytical result in \cite{Yang2014}, which showed
that \PT-symmetric potentials could not support continuous families
of non-\PT-symmetric solitary waves.

Numerically, we have found that these on-site and off-site bound
states are \emph{not} true solitary-wave solutions of equation
(\ref{Eq:psi}), in agreement with \cite{Yang2014} and contradicting
the predictions of equation \eqref{Eq:Centers}. Our numerical
verification used Newton-conjugate-gradient iterations
\cite{Yang_SIAM}, together with a multi-precision toolbox for Matlab
to resolve the numerical solutions to an accuracy of $10^{-18}$.
While there are approximate on-site and off-site bound-state
solutions, when investigated using multi-precision (24-digit)
calculations, we find that although the residue error can be as
small on the order of $10^{-13}$ they are not true solutions. This
disagreement between the exponential asymptotics and numerics is a
curious phenomenon. Based on this extremely small magnitude of the
residue error associated with these approximate solutions, we argue
that the non-existence of on-site and off-site bound states in the
\PT-symmetric periodic potential is due to terms which are smaller
than the leading-order exponentially small terms determined here.

\section{Conclusion}
Solitons in one-dimensional \PT-symmetric periodic potentials have
been studied using exponential asymptotics. Compared to previous
exponential asymptotics for conservative periodic potentials
\cite{Hwang2011,Akylas2012,Hwang2012}, the new feature of the
present analysis is that the inner and outer integral equations,
(\ref{e:Phi1})-(\ref{e:Phi2}) and (\ref{e:Outer1})-(\ref{e:Outer2}),
are both coupled systems due to complex-valued solitons.
Nonetheless, these coupled equations can be solved, thus the
exponential asymptotics analysis can be carried out. Following this
analysis, we show that two soliton families bifurcate out from each
Bloch-band edge for either self-focusing or self-defocusing
nonlinearity. An asymptotic expression for the eigenvalues
associated with the linear stability of these soliton families is
also derived. This formula shows that one of these two soliton
families (the off-site family) near band edges is always unstable,
while the other (on-site) family can be stable. In addition,
infinite families of two-soliton \PT-symmetric bound states,
comprising two on-site or two off-site solitons, have been
constructed by matching the exponentially small tails from two
neighboring solitons. These analytical predictions were compared
with numerics, and good overall agreements were observed. One minor
difference is that, while the exponential asymptotics analysis
predicts the existence of on-site and off-site two-soliton bound
states (which are non-\PT-symmetric), numerical computations
disprove their existence (see also \cite{Yang2014}). Based on the
numerical observation that the residue error of those approximate
on-site and off-site bound states is extremely small, we argue that
the non-existence of true on-site and off-site bound states is due
to terms which are smaller than the leading-order exponentially
small terms determined in this article.

\section*{Acknowledgment}
This work was supported in part by the Air Force Office of
Scientific Research (Grant USAF 9550-12-1-0244) and the National
Science Foundation (Grant DMS-1311730).

\end{document}